\documentclass[10pt, conference]{IEEEtran}

\usepackage[flushmargin,hang]{footmisc}
\renewcommand\footnoterule{%
  \kern-3pt\hrule width \columnwidth\kern2.6pt}
  
\usepackage{graphicx} 
\usepackage{cite}
\usepackage{amsmath,amssymb,amsfonts}
\usepackage{algorithmic}
\usepackage{graphicx}
\usepackage{textcomp}
\usepackage{xcolor}
\usepackage{multirow}
\usepackage{booktabs}   
\usepackage{tabularx}  
\usepackage{xurl}
\usepackage{amssymb}
\usepackage[hidelinks,breaklinks]{hyperref}
\title{TraceRAG: A LLM-Based Framework for Explainable Android Malware Detection and Behavior Analysis\\
}

\author{
\makebox[.33\linewidth][c]{\begin{tabular}{@{}c@{}}
Guangyu Zhang \\
\textit{Independent Researcher} \\
zgy020725@outlook.com
\end{tabular}}
\makebox[.33\linewidth][c]{\begin{tabular}{@{}c@{}}
Xixuan Wang \\
\textit{Australian National University} \\
xixuan.wang@anu.edu.au
\end{tabular}}
\makebox[.33\linewidth][c]{\begin{tabular}{@{}c@{}}
Shiyu Sun \\
\textit{George Mason University} \\
ssun20@gmu.edu
\end{tabular}}
\\[1.2em] 
\makebox[.33\linewidth][c]{\begin{tabular}{@{}c@{}}
Peiyan Xiao \\
\textit{The College of William and Mary} \\
pxiao@wm.edu
\end{tabular}}
\makebox[.33\linewidth][c]{\begin{tabular}{@{}c@{}}
Kun Sun \\
\textit{George Mason University} \\
ksun3@gmu.edu
\end{tabular}}
\makebox[.33\linewidth][c]{\begin{tabular}{@{}c@{}}
Yanhai Xiong\textsuperscript{*} \\
\textit{The College of William and Mary} \\
yxiong05@wm.edu
\end{tabular}}
}

\begin{document}

\maketitle
\begingroup
\renewcommand\thefootnote{}\footnotetext{\textsuperscript{*}Corresponding author.}%
\endgroup

\begin{abstract}
Sophisticated evasion tactics in malicious Android applications, combined with their intricate behavioral semantics, enable attackers to conceal malicious logic within legitimate functions, underscoring the critical need for robust and in-depth analysis frameworks. However, traditional analysis techniques often fail to recover deeply hidden behaviors or provide human-readable justifications for their decisions. Inspired by advances in large language models (LLMs), we introduce TraceRAG, a retrieval-augmented generation (RAG) framework that bridges natural language queries and Java code to deliver explainable malware detection and analysis. First, TraceRAG generates summaries of method-level code snippets, which are indexed in a vector database. At query time, behavior-focused questions retrieve the most semantically relevant snippets for deeper inspection. Finally, based on the multi-turn analysis results, TraceRAG produces human-readable reports that present the identified malicious behaviors and their corresponding code implementations. Experimental results demonstrate that our method achieves 96\% malware detection accuracy and 83.81\% behavior identification accuracy based on updated VirusTotal (VT) scans and manual verification. Furthermore, expert evaluation confirms the practical utility of the reports generated by TraceRAG.
\end{abstract}

\begin{IEEEkeywords}
Android Malware Detection, Malicious Behavior Analysis, Large Language Model, Retrieval-Augmented Generation
\end{IEEEkeywords}
\section{Introduction}
The rapid growth of mobile services and internet penetration has enabled individuals to engage in various activities through mobile applications, such as shopping, banking, and social networking. According to Statista's report \cite{statista}, the number of mobile app downloads worldwide has steadily increased from 2016 to 2023, when it reached 257 billion, and this upward trend is expected to continue. However, this surge in app usage has also made smartphones prime targets for cybercriminals, with Android devices being particularly vulnerable. The open-source nature of Android allows users to install apps from untrusted third-party markets, significantly increasing the risk of malicious software. In Q3 2024, the Kaspersky Security Network reported detecting 222,444 Android malware samples and potentially unwanted app variants \cite{Securelist}. Mobile malware spreads rapidly, and new variants of malicious apps emerge daily, often in millions \cite{Zimperium}. Cyber attackers are employing increasingly sophisticated techniques, such as obfuscation, sandbox evasion, and encryption, to evade detection. The proliferation of malware poses serious security threats, not only to individuals but also to businesses and government organizations, as it has the potential to compromise sensitive data, lead to financial losses, disrupt system functionality, and even enable large-scale cyberattacks. Consequently, developing effective methods for malware detection remains an urgent and critical issue. 

Researchers and practitioners have proposed three major categories of techniques to address such threats: static \cite{zhang2019efficient, lei2019evedroid, alazab2020intelligent, chen2024android}, dynamic \cite{feng2018novel, de2021detecting, surendran2020gsdroid}, and hybrid analysis methods \cite{han2019maldae, wang2022you, taher2023droiddetectmw, wu2023deepcatra, feng2025hgdetector}. Static analysis enables fast, scalable, and pre-installation malware detection by attempting to conservatively examine all possible execution paths \cite{samhi2024call}, but struggles with handling code obfuscation and dynamic runtime mechanisms such as virtual function calls, reflection, and event-driven execution—common techniques in modern mobile applications \cite{memon2015colluding, pan2020systematic, amin2020static}. In contrast, dynamic analysis does not rely on disassembly but instead directly observes an executable’s behavior at runtime, making it more effective against code obfuscation \cite{or2019dynamic, li2022novel}. However, it is time-intensive and resource-consuming \cite{ye2017survey}. Hybrid analysis combines both approaches to surmount their intrinsic limitations: it first performs static analysis and then observes the program dynamically at runtime, but faces the challenge of how to utilize the results of static analysis to assist dynamic analysis testing \cite{darabian2020detecting, baek2021two}. Additionally, most existing methods focus on detection or classification but lack interpretability, producing outputs that are neither human-readable nor insightful for thorough analysis. This raises reliability concerns, especially when applied to real-world or novel datasets, where their ability to detect previously unseen malicious Android applications remains questionable.

The rapid advancement of large language models (LLMs), exemplified by ChatGPT \cite{ChatGPT} and Llama \cite{touvron2023llama}, has transformed various fields. By leveraging vast datasets and advanced neural architectures, these models excel in language comprehension and generation, pushing the boundaries of artificial general intelligence and enabling effective collaboration with domain experts \cite{wei2022emergent, kaur2024text}. Building on these strengths, some researchers have explored the application of LLMs to enhance malware detection accuracy, such as analyzing behavioral semantics \cite{zhao2025apppoet}, capturing structural dependencies \cite{qian2025lamd}, enhancing interpretability through LLMs' non-decisional role \cite{li2024enhancing}, or leveraging retrieval-augmented generation (RAG) to transform static features into semantically rich functional summaries \cite{arikkat2025enhancing}. In contrast to these work leveraging various features as input, our approach directly provide LLM with Android Java source code, which provides a more comprehensive view of an app’s functionality and logic to uncover hidden malicious behaviors. Walton et al. \cite{walton2024exploring} introduce a hierarchical-tiered approach to code summarization, however, their method struggles with fully utilizing function call relationships, limiting its analytical precision. Our framework overcomes this limitation by generating summaries for code segments and producing credible analysis reports grounded in real code and call chains.

In this study, we present TraceRAG, an LLM-assisted system designed to analyze how Android malware carries out its malicious behavior. First, we construct a vector database to support a RAG pipeline. For each code segment, we employ a specific LLM to generate a descriptive summary, which is stored alongside the corresponding code. Second, we retrieve potentially suspicious code snippets from the database using a series of carefully crafted behavior-focused queries. Third, we prompt another specialized LLM with these retrieved snippets to perform a depth-first analysis of the code, guided by a structured and security-oriented prompt design. Finally, an additional LLM module compiles the analysis into a concise, human-readable report that includes app metadata, a summary of potential malicious behaviors, a step-by-step explanation of suspicious code paths and method calls, and an overall assessment conclusion. 

During the implementation of the proposed framework, a primary challenge lies in ensuring the precision of code snippets' retrieval for subsequent analysis. Given that a typical Android application may contain thousands of source files with complex structures and intricate interdependencies, identifying the most relevant code segments is not easy. To address this challenge, we carefully design prompts that enable the specific LLM to generate high-quality code summaries, which serve as semantic indices for RAG. Additionally, we store class and method names as metadata filters alongside the original code and corresponding summaries in the vector database, which further enhances retrieval accuracy.

Another significant challenge is the hallucination issue inherent in LLMs \cite{patsakis2024assessing}, which arises particularly when handling excessively long inputs or overly complex tasks, potentially resulting in incorrect or fabricated outputs \cite{zhang2023siren}. To mitigate this issue, we split the Java code into method-level chunks to reduce input length and remove dead code or unreachable statements to ensure that all the inputs fed into LLM are concise. In addition, we structure the analytical pipeline to support incremental LLM analysis by decomposing the broader task into smaller, clearly defined sub-tasks. Each sub-task is handled by a dedicated LLM, which reduces task complexity and cognitive load, thereby minimizing the likelihood of hallucinations. The detailed implementation of this stepwise approach is elaborated in the methodology section.

To assess TraceRAG’s performance, we conduct a comprehensive experimental study. First, we assemble an evaluation dataset of 70 malicious and 30 benign APKs from AndroZoo \cite{allix2016androzoo}. Using the original AndroZoo’s labels of datasets, TraceRAG achieves 90\% accuracy on binary malware detection and 83.81\% accuracy on behavior identification. After updating ground truth with recent VirusTotal (VT) scans and manual verification (involving source code analysis of suspicious behaviors and expert consultation), its malware detection accuracy rises to 96\%. We then compare TraceRAG’s reports against VT’s sandbox outputs, demonstrating our framework’s superior coverage, traceability, and behavioral organization. In addition, ablation experiments confirm the effectiveness of each retrieval enhancement technique in our framework. Finally, we deliver some generated analysis reports on malicious samples to experts from the Google Android Security Team for assessment to demonstrate the practicality and usefulness of our system.

In summary, we make the following contributions:

\begin{itemize}
\item[$\bullet$] To the best of our knowledge, our work is an initial exploration of applying RAG and LLM methods for Android malicious behavior detection, moving beyond traditional black‐box approaches to achieve code‐grounded analysis, and emphasizing the importance of explainability in parallel with decision-making.
\item[$\bullet$] We present TraceRAG, a framework that not only determines whether an application exhibits a specific malicious behavior but also pinpoints the exact Java code snippets and call chains responsible for it, offering clear advantages over existing malware analysis platforms.
\item[$\bullet$] Experimental results demonstrate that TraceRAG achieves high accuracy in both malware detection and behavior identification. Furthermore, we validate the quality and practical usefulness of its analysis reports through expert feedback. Our code and the generated reports  are available at \url{https://github.com/yanhaixiong/TraceRAG}.

\end{itemize}

\section{Literature review}
\subsection{Large Language Models in Code Understanding}
LLMs have demonstrated considerable potential in the domains of natural language understanding and programming code processing tasks \cite{fang2024large}. In the context of code understanding, LLMs can serve as on-demand information support by generating comprehensive explanations, detailed API descriptions, clarifications of domain-specific concepts, and illustrative usage examples \cite{nam2024using}. Moreover, LLMs can directly produce or facilitate the generation of executable and highly readable source code from decompiled binaries \cite{tan2024llm4decompile}, as well as identify and rectify errors within obfuscated disassembled code \cite{rong2024disassembling}, thereby improving reverse engineering performance. To further enhance development efficiency and improve code maintainability, recent studies have employed methods such as few-shot training for project-specific adaptation and semantic prompt augmentation to enhance the performance of LLMs in code summarization tasks \cite{ahmed2024automatic, ahmed2022few}. In malware summarization, Lu et al. \cite{lu2024malsight} fine-tune CodeT5+model using transfer learning, integrating malicious software functional features and decompiled pseudocode structural characteristics to generate informative code summaries. Furthermore, Walton et al. \cite{walton2024exploring} leverage OpenAI’s GPT-4o-mini model with optimized prompt engineering to automatically classify Android malware and generate functional summaries at function, class, and package levels, thereby enabling systematic security analysis and tracing of malicious behaviors.

\subsection{Large Language Models in Malware Analysis}
Conventional malware detection methods often struggle against sophisticated and polymorphic malware designed to evade detection \cite{al2024exploring}. Advancements in LLMs introduce novel methodologies that overcome these limitations by leveraging extensive pre-trained knowledge to identify subtle coding patterns in malware \cite{jelodar2025large}. Some researchers have applied LLMs to perform static analysis on Android apps, facilitating effective malware classification and generating detailed explanatory insights \cite{rahali2021malbert, zhao2025apppoet}. For instance, Qian et al. \cite{qian2025lamd} construct a practical, context-driven framework that employs static analysis combined with backward program slicing to extract sensitive API contexts, followed by a three-tier LLM reasoning pipeline enhanced with factual consistency verification, improving the accuracy of malware detection. Additionally, Li et al. \cite{li2024enhancing} compare traditional decision-centric Android malware detection models with ChatGPT, demonstrating the superiority of LLMs’ non-decisional contributions in providing detailed analysis. Similarly, Arikkat et al. \cite{arikkat2025enhancing} propose a RAG framework that transforms structural Android app features into semantically rich descriptions, achieving superior classification accuracy when combined with domain-specific BERT classifiers. Furthermore, employing a hybrid testing approach, Wang et al. \cite{wang2024liredroid} integrate static analysis of API call chains with LLM-enhanced test case generation and dynamic code injection to replicate and detect sensitive behaviors. The application of LLMs for malware detection also extends beyond Android platforms into diverse environments and data formats, including websites \cite{koide2024chatphishdetector}, Windows systems \cite{devadiga2023gleam, li2023efficient, wang2024unmasking, zhou2025srdc}, Java source files \cite{hossain2024malicious, shestov2025finetuning} and NPM Packages \cite{yu2024maltracker, huang2024spiderscan}.

While existing approaches have advanced Android malware detection through various learning-based methods, they fundamentally operate at the abstraction level of extracted features. In contrast, our approach directly analyzes Java source code at the method level, establishing semantic bridges between natural language queries and actual code implementations through RAG. This shift from feature-level to code-level analysis enables not only detection but also precise localization of malicious logic within applications, providing security analysts with traceable evidence paths from suspicious behaviors to their concrete implementations. 
\section{Proposed Methodology}
\begin{figure*}[ht]
    \centering
    \includegraphics[width=1.0\textwidth]{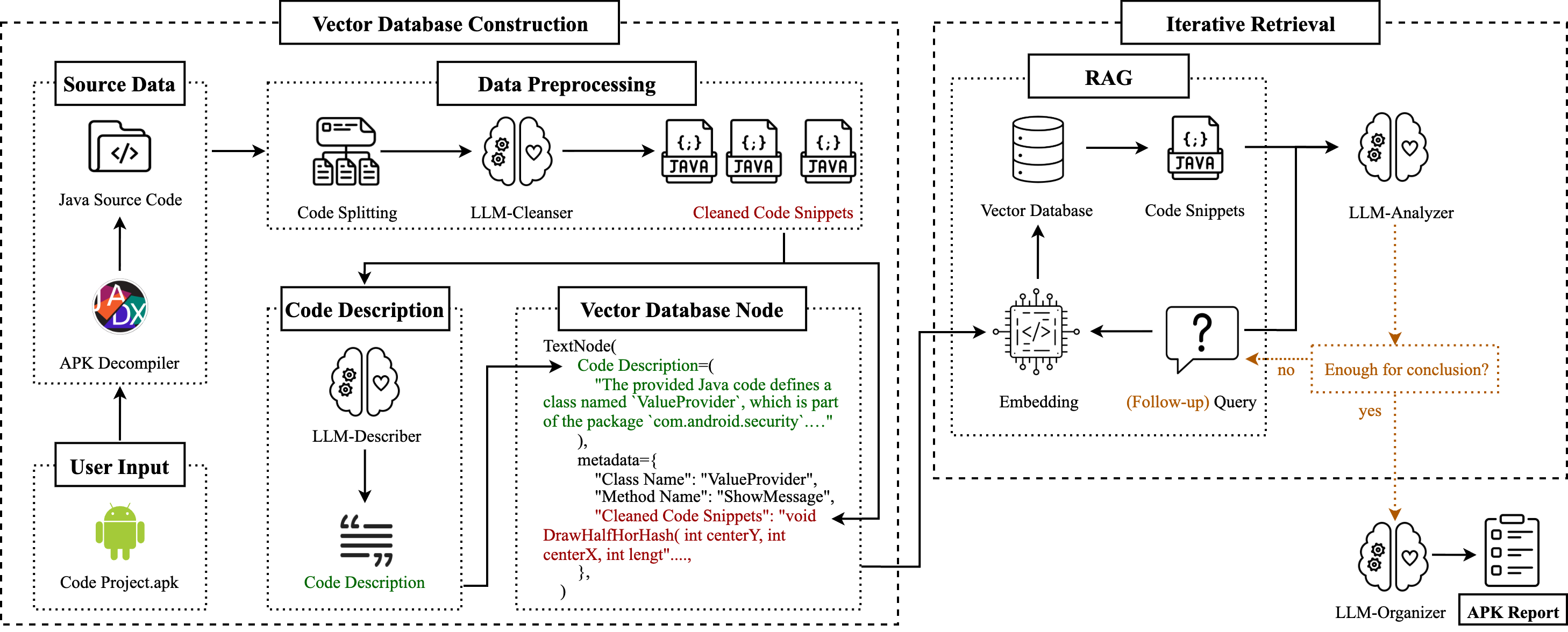}
    \caption{System Architecture of TraceRAG}
    \label{architecture}
\end{figure*}

This section presents an overview of TraceRAG. The system begins by reverse-engineering an Android application to extract all associated Java files. These files are then segmented and cleaned by a dedicated LLM, named LLM-Cleanser, to improve the quality of subsequent analysis. For each processed code segment, a textual description is generated by LLM-Describer and used as the index, while both the description and the corresponding code are stored together in the vector database for retrieval. Subsequently, relevant code segments are retrieved from the database based on carefully designed queries and analyzed by LLM-Analyzer to detect potential malicious behaviors within the app. Based on the LLM-Analyzer’s iterative analysis, the system generates a report summarizing any suspicious activities detected and provides a clear, comprehensive explanation. The overall system architecture of TraceRAG is shown in Fig.~\ref{architecture}. For the call chain, the output of one step serves as the input to the next, and we implement this agent-like sequence using a LangGraph workflow.

\subsection{RAG Framework}
In this study, we employ a RAG framework combined with LLMs to facilitate malicious software detection in Android applications. 

In a typical RAG application, there are two primary components: (1) an indexing pipeline for ingesting data from a source and constructing an index, and (2) a retrieval-and-generation mechanism that processes a user query in real time, retrieves relevant data from the index, and passes it to the model \cite{lewis2020retrieval}. However, our framework differs in a crucial respect: rather than retrieving text-based content, our goal is to retrieve Java code. Traditional RAG systems assess the semantic similarity between text queries and text documents, but Java code—lacking natural language semantics—cannot be directly retrieved in this manner.

To overcome this challenge, we leverage LLM-Describer to generate descriptive summaries for each Java code, articulating its functionality and potential usage in human-readable text. These generated descriptions serve as “indexes” that enable retrieval based on semantic similarity with the user’s natural-language query. In practical terms, we store each Java code snippet alongside its corresponding description and metadata in a vector database. Consequently, when the system receives a query in human language, it compares the query with the code’s textual descriptions and accurately retrieves the pertinent code segments, thus fulfilling the primary goal of our RAG-based framework. Additionally, we create a separate vector database for each app, to ensure that no Java code from other apps is mixed in and to avoid collisions.

We adopt the RAG paradigm for three primary reasons. First, an Android application inherently serves as a database, thus establishing a well-defined scope for building a vector database. In addition, the Java code within the app naturally consists of distinct code segments, effectively mitigating the challenges associated with chunking. Moreover, despite Java code, as a machine-oriented language, lacks inherent human-language semantics, recent advancements in LLMs enable accurate code understanding and summarization. Consequently, these capabilities enable precise retrieval of Java code blocks that semantically align with human-language queries, ensuring that only relevant code is selected for analysis.

All LLM modules in our framework (LLM-Cleanser, LLM-Describer, LLM-Analyzer, LLM-Organizer and several Reviewers) utilize the OpenAI o3-mini model \cite{OpenAI}, a compact model optimized for reasoning and code analysis. Its low computational cost and strong performance make it particularly well-suited for our task within the TraceRAG pipeline.

\subsection{Description Generation}

In this section, we present our method for generating high-quality descriptions of Java code. This process includes APK decompilation, code splitting and cleaning, as well as description generation, which collectively support accurate and efficient semantic retrieval in downstream analysis.

\subsubsection{APK Decompilation}
We begin by decompiling the Android APK using the reverse engineering tool JADX \footnote{\url{https://github.com/skylot/jadx}} to obtain all associated Java source files. To enhance the clarity and traceability of the final report, we also extract key metadata from the decompiled content, including package name and SHA-256.

\subsubsection{Code Splitting and Cleaning}
A major challenge at this stage is that some Java files contain millions of tokens, far exceeding the input limits of LLMs and degrading the quality of generated descriptions. Given that Java code follows a well-defined structure—where classes encapsulate methods—we apply method-level code splitting using the JavaLang library in Python. Each method is extracted individually, while preserving essential information such as imported packages and class-level variable declarations. In the end, we create each of the split methods a file, converting a long class level file into many smaller method level files.

\begin{figure*}[ht]
    \centering
    \includegraphics[width=0.95\textwidth]{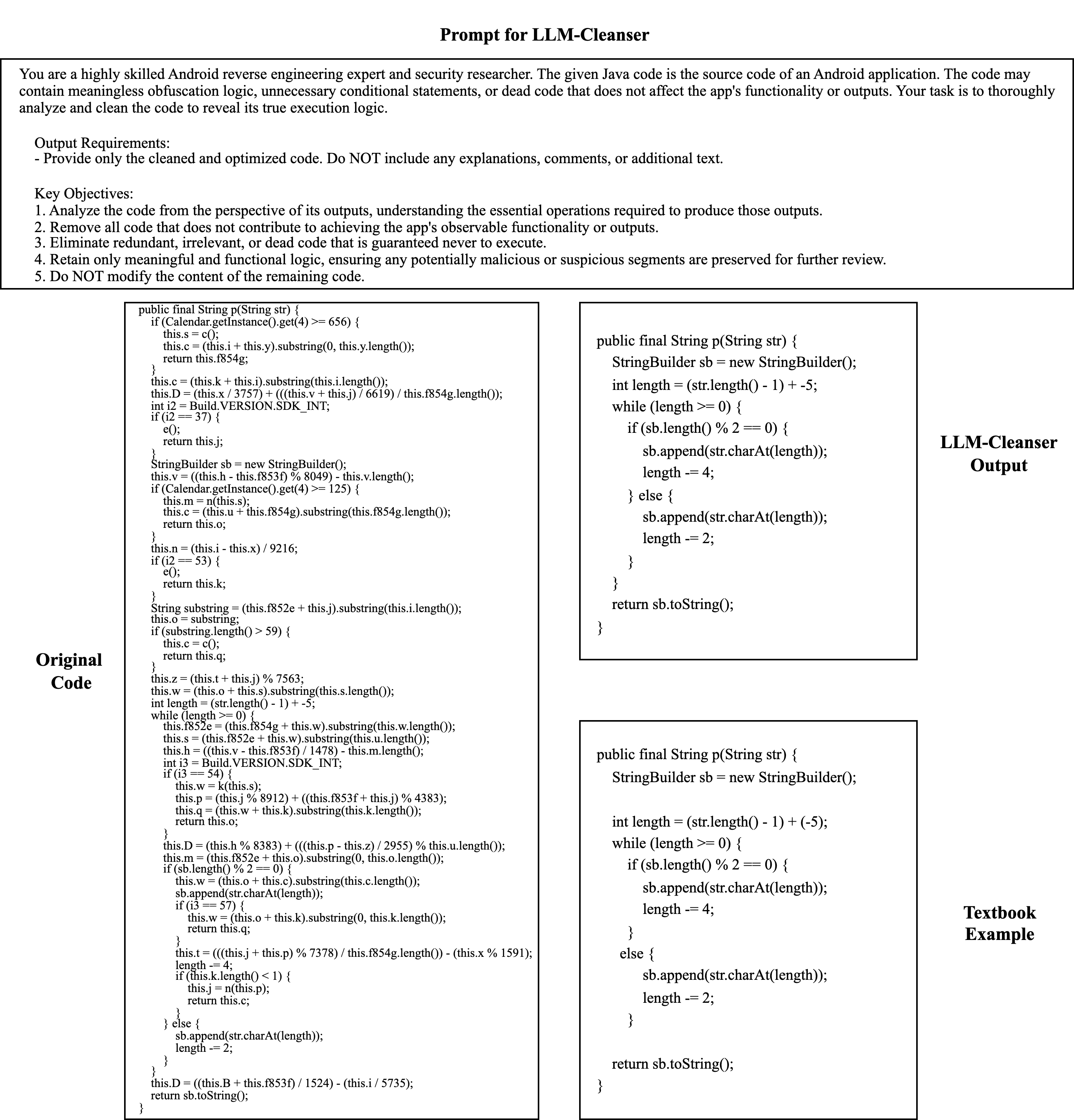}
    \caption{Prompt and Cleaning Case for LLM-Cleanser}
    \label{clean}
\end{figure*}

Another factor affecting description quality is code obfuscation, which is common in real-world Android applications. Obfuscated code often contains dead code and unreachable branches that do not contribute to actual execution. To mitigate this, we use LLM-Cleanser to remove such irrelevant code, preserving only the core logic. This results in clearly organized code snippets of manageable length, making them well-suited for the next step of description generation. Fig.~\ref{clean} presents the exact prompt used by LLM-Cleanser together with a representative cleaning case on the classic Android malware \texttt{com.bp.statis.bloodsugar}. The cleaned snippet clearly eliminates unreachable and opaque code while preserving the essential semantics, resulting in a structure closely resembling the textbook implementation. This similarity demonstrates the effectiveness of our cleaning step.

\subsubsection{Code Description}

\begin{figure*}[ht]
    \centering
    \includegraphics[width=1.0\textwidth]{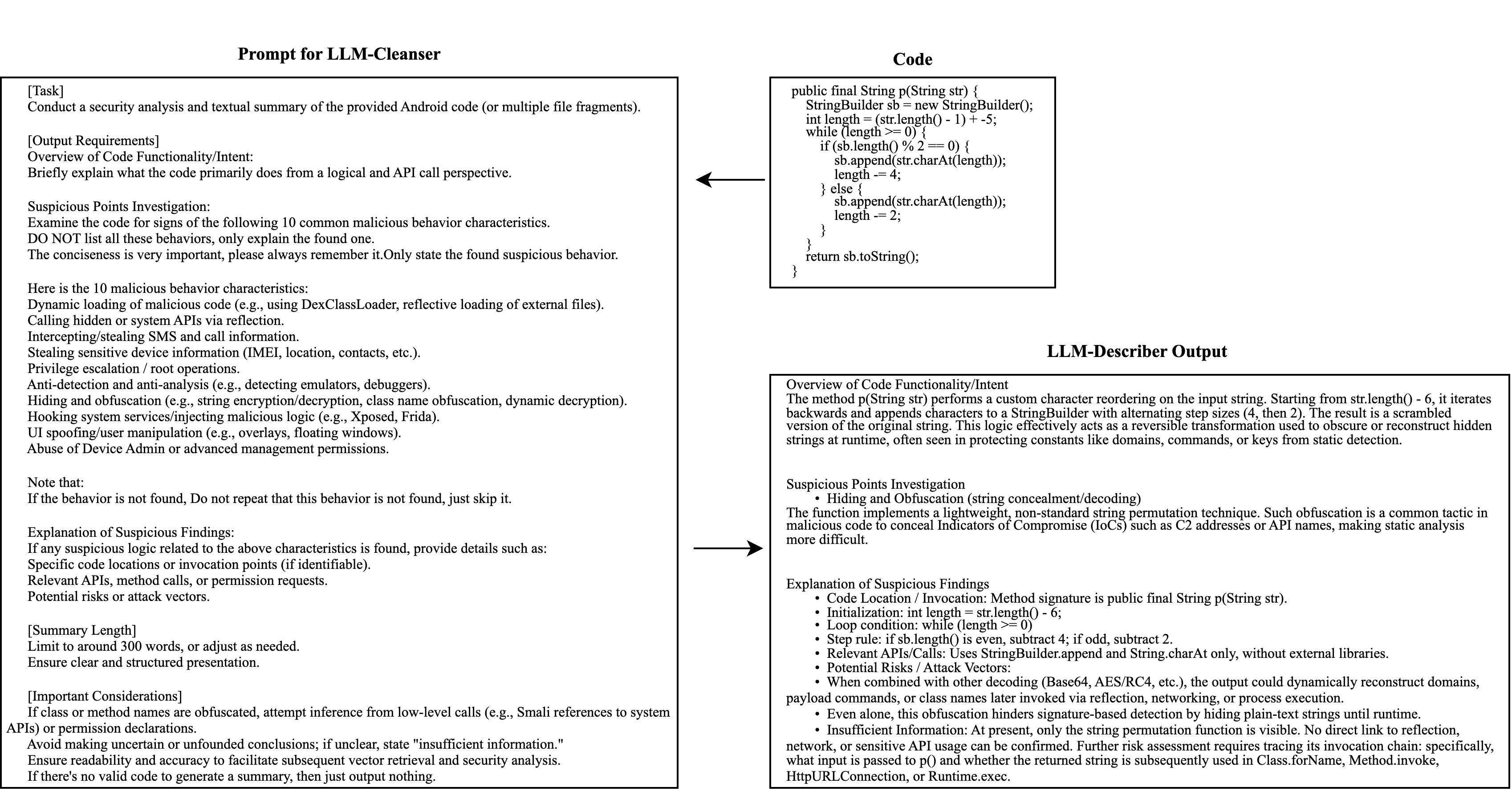}
    \caption{Prompt and Description Case for LLM-Describer}
    \label{description}
\end{figure*}

We leverage LLM-Describer to generate descriptions of code snippets through carefully designed prompt engineering. The structure of our prompt template consists of two key components: First, the LLM-Describer is instructed to focus on the core functionality of the code—explaining in detail what the code does, including its inputs and outputs. Second, the prompt guides the LLM-Describer to identify and describe any potentially malicious intent. The output includes all observed suspicious behaviors, enabling more accurate semantic retrieval when using queries related to specific threats. Fig.~\ref{description} presents the exact prompt used by LLM-Describer together with a representative case on the \texttt{p(String)} method from \texttt{com.bp.statis.bloodsugar}. The output successfully identifies the string permutation obfuscation and highlights suspicious behaviors, which aligns with our expected objectives for the description step.

\subsection{Vector Database Construction}

To improve the accuracy of code retrieval, we not only index the LLM-generated code descriptions but also incorporate metadata as auxiliary labels to help categorize and filter code snippets. Specifically, we include the following elements for each record, enabling precise filtering when a query requires a specific method definition:

\begin{itemize}
\item[$\bullet$] Processed Code Snippets: The method-level code snippets extracted from the application after splitting and cleaning.
\item[$\bullet$] Code Description: A summary generated by the LLM describing the method’s functionality and potential malicious behavior.
\item[$\bullet$] Method Name: The name of the method.
\item[$\bullet$] Class Name: The name of the class to which the methods belong.
\end{itemize}

We employ OpenAI's text-embedding-ada-002 model to generate embeddings for each record \cite{OpenAIembedding}, encompassing both the description index and associated metadata.

\subsection{LLM Conversation}
With all preparatory steps completed, we proceed to analyze the suspicious behaviors of the target application through interactions with LLM-Analyzer.
 
\subsubsection{Query Design}
To conduct a comprehensive analysis, we design 11 retrieval queries covering three common categories of malicious behavior observed in Android malware, as shown in Table~\ref{query}. These queries are formulated based on empirical experiments and expert input from professional malware analysts. Additionally, they are intentionally kept simple for ease of interpretation and to align better with LLM reasoning, which reduces ambiguity in multi-turn analysis. Additionally, the query set can be easily updated to reflect emerging malware patterns, supporting incremental indexing of recent apps and behaviors. For each APK under inspection, all 11 queries are executed sequentially, with each query invoking the complete iterative retrieval pipeline before proceeding to the next, thereby ensuring full coverage of all targeted behaviors. 

\begin{table}[!htbp]
  \centering
  \caption{Query Design for TraceRAG}
  \label{query}
  \begin{tabularx}{\columnwidth}{@{}>{\raggedright\arraybackslash}p{2.5cm}|X@{}}
    \toprule
    \textbf{Type} & \textbf{Query} \\
    \midrule
    \multirow[t]{4}{2.5cm}{Information Theft and Abuse}
      & Q1: Does the application access or collect sensitive user data (e.g., SMS, contacts, location, or device identifiers)? \\
      & Q2: Does the application capture user activity through screen recording or screenshots? \\
      & Q3: Does the application connect to suspicious external URLs or perform background downloads without user interaction? \\
      & Q4: Is obfuscation or encryption used to conceal communication endpoints or downloaded content? \\
    \cmidrule{1-2}

    \multirow[t]{3}{2.5cm}{Monetary Fraud and Financial Abuse}
      & Q5: Does the application send messages or make calls that may incur charges without user consent? \\
      & Q6: Does the UI mislead users into clicking ads or subscribing to services? \\
      & Q7: Is there evidence of tampering with in-app purchases or payment processes? \\
    \cmidrule{1-2}

    \multirow[t]{4}{2.5cm}{Privilege Abuse and System Exploitation}
      & Q8: Does the application request elevated privileges (e.g., Accessibility or Device Administrator) or attempt to maintain persistence? \\
      & Q9: Does the application support remote command execution or include dynamic code loading and anti-analysis techniques? \\
      & Q10: Is there evidence of root-level activity, such as executing system commands or interacting with system partitions? \\
      & Q11: Does the application use native libraries or known exploits to escalate privileges or bypass system security policies? \\
    \bottomrule
  \end{tabularx}
\end{table}

\subsubsection{Retrieval and Result Processing}
A critical aspect of retrieval is setting an appropriate threshold. A overly high threshold may result in few or no matches, potentially excluding truly malicious code. Conversely, a low threshold may yield many irrelevant results, increasing analysis time and degrading the quality of the final report. To strike a balance between stability and relevance, we employ a two-stage retrieval strategy. First, we apply a top-k threshold of five to ensure a consistent number of code snippets retrieved. Then, our Relevance-Reviewer LLM filters out irrelevant snippets and retains only those indicative of suspicious behavior. If none of the top-5 retrieved snippets are relevant to the query or demonstrate any indicators of malicious behavior, this LLM outputs a message that indicates no related code is found, and the query result is excluded from further analysis. This approach reduces noise while maintaining retrieval consistency, thereby improving the accuracy and efficiency of downstream processing.

\subsubsection{LLM Analysis}
Following retrieval and initial filtering by Relevance-Reviewer, the remaining code snippets—those most likely to contain malicious behavior—are analyzed individually by LLM-Analyzer to determine whether they exhibit malicious intent. The analysis proceeds in several steps. First, LLM-Analyzer identifies the core behavior of the given code, determines its intent, and reports the fully qualified code path. If the code does not exhibit malicious behavior, or if a conclusion can already be drawn based on the current snippet, LLM-Analyzer will generate a detailed result summarizing the identified behavior and include the corresponding code path for reference.

If LLM-Analyzer determines that the current code snippet invokes another function to achieve a specific intention, and additional context is required to understand its behavior and reach a conclusion, it will generate a follow-up query. This query includes the target method’s name, the corresponding class name, and its input parameters to support further analysis. For example: 

\textit{“Could you provide the implementation of the method j defined in class b, which takes one String as an input parameter?”}

This newly generated query is then used to retrieve from the vector database mentioned before. During retrieval, the stored metadata is applied to narrow the search scope. In this example, the system filters for code snippets labeled with \textit{"method name j"} and \textit{"class name b"}. Typically, this yields a single match; however, to address potential naming collisions, all retrieved snippets are passed to Collision-Reviewer. This LLM-based reviewer selects the most appropriate result, which is then forwarded to LLM-Analyzer for continued analysis. We also implement Query-Reviewer that inspects LLM-Analyzer’s output and, if it detects a follow-up query, invokes the retrieval module; otherwise, it generates the final results. 

\begin{figure*}[ht]
    \centering
    \includegraphics[width=1.0\textwidth]{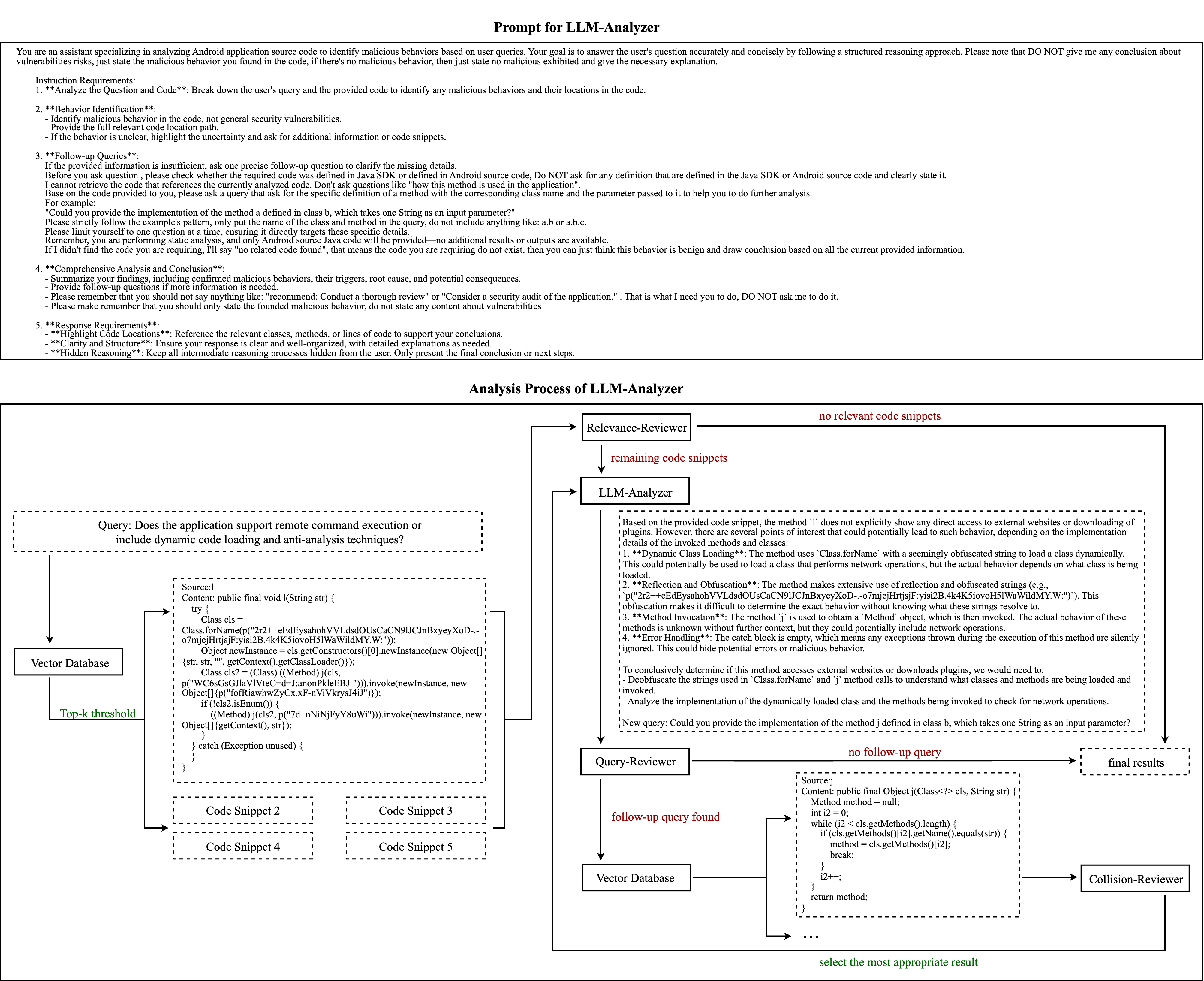}
    \caption{Prompt and Analysis Process of LLM-Analyzer}
    \label{process}
\end{figure*}

It is worth noting that although both LLM-Describer and the LLM-Analyzer are involved in interpreting code behavior, they serve distinct purposes. LLM-Describer only focuses on summarizing the general intent of a given code snippet. Its primary task is to clearly describe the code’s functionality, including its inputs, outputs, and overall purpose. In contrast, LLM-Analyzer is designed to conduct a deeper investigation into the code’s usage context, with the goal of identifying malicious behavior by digging out all components contributing to the malicious functionality. Fig.~\ref{process} illustrates the prompt and a segment of the LLM-Analyzer’s workflow on \texttt{com.bp.statis.bloodsugar}. The results indicate that the LLM-Analyzer not only pinpoints suspicious behaviors but also formulates appropriate follow-up queries, thereby validating the practicality of our analysis step.

\subsubsection{Report Generation}

The reporting process for results generated by LLM-Analyzer is organized into three hierarchical layers: code reports, query reports, and a final report. Each layer is produced by a dedicated LLM-Organizer. First, each code snippet may require multiple rounds of analysis and supporting code retrieval. An LLM-Organizer compiles these results into a code report. Then, all code reports under the same query are aggregated into a query report. Finally, all eleven query reports are combined into a final report. This final report is structured into four main parts: App Info, Overall Summary, Detailed Analyses, and Conclusion. In the Detailed Analyses section, each query’s subsection either provides a detailed summary of the identified malicious behavior, complete with implementation details and supporting code references, or explicitly states that no related malicious activity is detected.

\section{Evaluation}
In this section, we evaluate the performance of TraceRAG by answering the following research questions (RQs):

\begin{itemize}
\item[$\bullet$] \textbf{RQ1}: How effective is TraceRAG in detecting Android malware and malicious behavior?
\item[$\bullet$] \textbf{RQ2}: Do the preprocessing steps of the RAG module enhance its effectiveness and robustness?
\item[$\bullet$] \textbf{RQ3}: Are the reports generated by TraceRAG instructive and valid?
\end{itemize}

\subsection{Datasets}
To rigorously assess TraceRAG’s performance, we use a dataset sourced from AndroZoo \cite{allix2016androzoo}, a continuously growing collection of Android applications gathered from multiple official App stores like Google Play. To ensure that our samples cover the vast majority of real-world APKs, we randomly download 1,000 apps from AndroZoo spanning 2010 to 2025, each of which has been scanned on VT by more than ten security scanners\footnote{\url{https://www.virustotal.com/gui/home/upload}}. After APK decompilation and code splitting described above, we find that most samples contain fewer than 3,000 code snippets. We therefore randomly select 100 apps within this range—30 benign and 70 malicious, based on AndroZoo’s labels—as our evaluation dataset. This ratio is used to expose TraceRAG to more malicious behaviors, since the goal is behavior analysis rather than binary classification. Table~\ref{data} shows details about the used dataset. Statistics show that generating the 100 corresponding analysis reports consumes over 100 million total tokens and incurs approximately \$600 in API charges.

\begin{table}[ht]
  \centering
  \caption{Overview of the Evaluation Dataset}
  \label{data}
  \begin{tabular}{l|c|c}
    \toprule
    \textbf{Attribute} &  \textbf{Interval Range} & \textbf{Sample Count} \\
    \midrule
    \multirow{3}{*}{Snippet Count Before Splitting}
       & $<100$            &  54  \\
       & 100--200         &   26 \\
       & 200-400           &  20 \\
    \midrule
    \multirow{3}{*}{Snippet Count After Splitting}
      & $<1000$            &  48  \\
      & 1000--2000        &   16 \\
      & 2000-4000           &  36  \\
    \midrule
    \multirow{3}{*}{Size (MB)}
        & $<5$             &  65  \\
        & 5--20            &   26 \\
        & 20--50             &  9  \\
    \bottomrule
  \end{tabular}
\end{table}

\subsection{RQ1: Performance of TraceRAG}
To answer RQ1, we assess TraceRAG across three related evaluations:

\begin{itemize}
\item[$\bullet$] \textbf{Evaluation 1}: A binary malware detection evaluation using AndroZoo labels as ground truth. 
\item[$\bullet$] \textbf{Evaluation 2}: A refined binary malware detection evaluation using VT updated results supplemented with manual verification as ground truth to account for discrepancies between AndroZoo labels and VT results.
\item[$\bullet$] \textbf{Evaluation 3}: A multiclass behavior detection evaluation in which each malicious app is assigned to one or more of the three predefined behavior categories (Table~\ref{query}), with predictions compared against VT’s behavior analysis results. 
\end{itemize}

The evaluation metrics employed in our experiments are accuracy, precision, recall, and F1‐score. For the multiclass evaluations, since each malicious APK may exhibit multiple behavior types, we treat each of the three behavior categories as a separate binary classification and compute micro-averaged precision, recall, and F1-score.

When evaluating TraceRAG on benign samples, we note that it occasionally reports potential vulnerabilities or risks rather than true malicious behaviors. This effect stems from our prompt’s instruction to flag any suspicious code, leading TraceRAG to treat security risks as indicators of malware. To prevent conflating non-malicious vulnerabilities with malware detection, we establish a criterion: a sample is counted as malicious only when the report explicitly identifies a malicious behavior and provides a corresponding explanation of its implementation. Generic warnings, such as “this code may pose risks when used”—are therefore not considered evidence of malware. By applying this rule, we ensure that TraceRAG’s performance metrics reflect genuine malware detection rather than general security concerns.

Table~\ref{performance} summarizes the results for the three evaluation settings described above, demonstrating that TraceRAG achieves strong overall performance. Specifically, in the first malware detection evaluation based on AndroZoo labels, TraceRAG obtains an accuracy of 90\%, correctly identifying all 70 malicious APKs (achieving a recall of 100\%), while misclassifying 10 benign samples as malware. For behavior detection, TraceRAG achieves an overall accuracy of 83.81\%, with precision, recall, and F1‐scores similarly high across the dataset. These results also demonstrate the effectiveness of each LLM-driven module in our pipeline.

\begin{table}[!t]
\centering
\caption{Overall Performance (\%) of TraceRAG on Malware and Behavior Detection}
\label{performance}
\begin{tabular}{l|c|c|c|c}
\toprule
\textbf{Setting} & \textbf{Accuracy} & \textbf{Precision} & \textbf{Recall} & \textbf{F1}  \\
\midrule
\textbf{Evaluation 1} & 90.00 & 87.50 & 100.00 & 93.33 \\
\midrule
\textbf{Evaluation 2} & \textbf{96.00} & \textbf{95.89} & \textbf{98.59} & \textbf{97.22} \\
\midrule
\textbf{Evaluation 3} & 83.81 & 84.89 & 86.35 & 85.46 \\
\bottomrule
\end{tabular}
\end{table}

\subsubsection{Malware Detection Performance}

To obtain more accurate performance estimates, we re‐examine the 30 samples originally labeled benign by AndroZoo, using updated VT scans and manual verification. Among the 10 false positives reported by TraceRAG, seven are found to be valid detections upon closer inspection. In addition, we identify one actual malicious sample that TraceRAG failed to detect—a false negative. Correcting these discrepancies raises our effective malware detection accuracy from 90\% to 96\%.

Among the ten false positive cases in Evaluation 1 that use AndroZoo labels as ground truth, one common scenario involves samples that are still labeled as benign by AndroZoo but have been reclassified as malicious in more recent VT scans. For example, in the case of \texttt{com.smartsm5.smart\_5\_293 (sha256: 2A88D86B5F36EFD0E9A668B84C893171C0A9326DD1\\54FAEF1A35F80884F5BED7)}\footnote{\url{https://www.virustotal.com/gui/file/2a88d86b5f36efd0e9a668b84c893171c0a9326dd154faef1a35f80884f5bed7}}, TraceRAG’s report includes the following excerpt:

\begin{quote}
Within the run() method, the app accesses an external URL (via main.this.downloadImgUrl) and downloads an image silently.

- The downloaded image is stored in the device’s external storage under the DCIM/Camera folder without any user notification.

- The call chain is clear: execution enters run() → retrieves the URL from main.this.downloadImgUrl → performs a silent download and writes to external storage, reflecting covert behavior.
\end{quote}

Although this sample is labeled as benign by AndroZoo (based on a VT scan dated 2014-06-26), more recent VT results have reclassified it as malicious (dated 2021-02-05), citing external URL-related suspicious behavior. Three other cases exhibit similar discrepancies between historical and updated VT results, further validating TraceRAG’s capability to detect malware behavior that may have been previously overlooked.

Apart from such outdated-labeling cases, we also observe examples where VT fails to detect actual malicious behavior. One such case is \texttt{com.dijlah.sh\_khotaba 
(sha256:55152EE88521E599145568F8CF949BAA4D\\9884B6670002C8FC760844CD540947)}\footnote{\url{https://www.virustotal.com/gui/file/55152ee88521e599145568f8cf949baa4d9884b6670002c8fc760844cd540947}}, which is also labeled benign in AndroZoo. However, TraceRAG's result shows that it allow the application to perform unauthorized financial operations by sending SMS and initiating phone calls without user confirmation. We manually confirm that this app’s code indeed contains silent SMS-sending functions, with permissions to modify both the message content and the recipient. Two additional samples fall into this category, where TraceRAG successfully identifies malicious functionality that is not flagged by VT.

The remaining three cases appear to be genuine false positives. In these instances, TraceRAG infers that the apps are attempting to connect to an external URL and silently conduct a download behavior without the user's consent. However, our inspection reveals that the apps are merely conducting regular connection functions without evidence of malicious intent. This suggests a limitation of LLM-based reasoning: lacking access to the actual content of the external URL, the model tends to conservatively classify ambiguous URL-related behavior as malicious. 

\subsubsection{Behavior Detection Performance}
To evaluate TraceRAG’s ability to detect each behavior type, Table \ref{classification} presents the per‐category precision, recall and F1‐score across all 70 malicious samples. In total, TraceRAG correctly identifies 176 of the 210 labeled behaviors. However, the results indicate that behaviors associated with \textit{Monetary Fraud and Financial Abuse} present greater challenges for accurate detection, with noticeably lower classification performance compared to the other categories. By analyzing TraceRAG’s reasoning process for this type, we identify three primary misclassification causes, each corresponding to one of the three queries under this category. First, sending SMS may be part of an app’s legitimate functionality (e.g., for verification via one-time passwords). However, if TraceRAG cannot reliably determine whether the appropriate permissions have been requested or whether user consent has been properly obtained, it may incorrectly classify such behavior as malicious. Second, assessing whether the user interface is misleading involves interpreting elements related to front-end design, button labels, and behavioral cues. Such interaction-related information is typically absent from the Java source code, which may lead to misinference by TraceRAG. Third, in-app purchases and payment processes often span multiple modules. Developers may also employ techniques such as reflection, dynamic loading, and code obfuscation to conceal critical operations. Even with some capacity for behavioral tracing, TraceRAG may fail to capture the full execution chain or detect subtle manipulations. As the invocation chain becomes longer and more complex, TraceRAG is more prone to misclassification.

\begin{table}[!t]
\centering
\caption{Detailed Performance (\%) of TraceRAG on Behavior Detection}
\label{classification}
\begin{tabular}{
    >{\raggedright\arraybackslash}m{2.6cm}|
    >{\centering\arraybackslash}m{1cm}|
    >{\centering\arraybackslash}m{1cm}|
    >{\centering\arraybackslash}m{1cm}|
    >{\centering\arraybackslash}m{1cm}
}
\toprule
\textbf{Type} & \textbf{Accuracy} & \textbf{Precision} & \textbf{Recall} & \textbf{F1}  \\
\midrule
\textbf{Information Theft and Abuse}
     & 88.57 &98.41  &89.86  &93.94 \\  
\midrule
\textbf{Monetary Fraud and Financial Abuse} 
      & 75.71 &68.75  &75.86  &72.13 \\
\midrule
\textbf{Privilege Abuse and System Exploitation}
      & 87.14 &87.50  &93.33  &90.32 \\
\bottomrule
\end{tabular}
\end{table}

\subsubsection{Comparison with VirusTotal Reports}

Currently, VT offers several sandbox analyses—Tencent HABO, Zenbox android, VirusTotal Droidy, and VirusTotal R2DBox, but each delivers either behavioral logs or source‐code snippets in isolation and lacks explicit linkage between observed behaviors and their implementing code. Moreover, their coverage remains low: on 100 randomly selected APKs, none exceed 40\% coverage (as shown in Table~\ref{report}). In contrast, TraceRAG can generate a comprehensive report for every APK: it not only identifies suspicious behaviors but also retrieves the exact Java snippets responsible and reconstructs the full call chains that realize those behaviors, achieving 100\% coverage in our evaluation. In addition, by clearly categorizing malicious behavior types and pinpointing their root causes, TraceRAG can help experts narrow their focus, significantly reducing investigation scope and saving valuable time.

\begin{table*}[!t]
\centering
\caption{Comparison of TraceRAG and VT Reports}
\label{report}
\begin{tabular}{l|c|c|c|c|c|c}
\toprule
 & \textbf{Behavior} & \textbf{Source code} & \textbf{Link between them} & \textbf{Coverage Ratio} & \textbf{Malicious}  & \textbf{Benign}\\
\midrule
\textbf{Tencent HABO} &\checkmark & $\times$  & $\times$ & 36\% &28 &8 \\
\midrule
\textbf{Zenbox android} &\checkmark & $\times$ & $\times$  &1\%  &1  & 0 \\
\midrule
\textbf{VirusTotal Droidy} &$\times$ & \checkmark & $\times$  & 37\%  &26 &11  \\
\midrule
\textbf{VirusTotal R2DBox} &$\times$ & \checkmark & $\times$  & 3\%  &3 &0  \\
\midrule
\textbf{TraceRAG} & \checkmark & \checkmark & \checkmark & 100\% &70 &30 \\
\bottomrule
\end{tabular}
\end{table*}

\subsection{RQ2: Ablation Study on TraceRAG}

Since TraceRAG's analysis depends critically on the quality of retrieval and the additional information provided during analysis, any change in these components can significantly affect the final report. To evaluate the necessity of each retrieval enhancement technique in our framework, we conduct ablation experiments from three perspectives: the role of code descriptions, the impact of preprocessing, and the influence of multi‐turn interaction. To maintain consistency, we continue to use \texttt{com.bp.statis.bloodsugar} as our example, which contains 9,107 code snippets before splitting and 86,035 snippets after splitting.


\subsubsection{Role of Code Description}
As described before, TraceRAG relies on code descriptions generated by LLM-Describer to index and retrieve the most relevant code snippets for each malicious behavior query, which are then passed to LLM-Analyzer for further analysis. To validate the importance of the description step, we perform an ablation experiment in which we skip description generation and instead index only the raw source code. In \texttt{com.bp.statis.bloodsugar} case study, indexing with code descriptions achieves successful retrievals for most queries. In contrast, indexing on raw code alone returns no valid matches among the top five results, making any downstream analysis impossible. 

These results further support one of this paper’s contributions: by leveraging LLMs and RAG, we establish a bridge between natural language queries and Android Java code, enabling simple user queries to retrieve the most relevant snippets. This approach is not limited to Java, it can be extended to other languages such as Python or C++ and explored in a variety of related research domains.

\subsubsection{Effect of Preprocessing}
Before the description generation process of TraceRAG, we first apply a series of preprocessing steps, including code splitting and cleaning. This is motivated by the observation that many Java files are excessively long and contain heavily obfuscated code, which can hinder code understanding and retrieval effectiveness. To evaluate the necessity of these preprocessing steps, we conduct an ablation in which we skip splitting and cleaning, and instead pass the entire original Java file to LLM-Describer for description generation. To account for this change, we also adjust the prompt for the LLM-Describer to handle entire Java file, which typically contain a single public class with multiple methods. We instruct the model to provide an overview of this class along with the explaination of each method's intent and functionality.

As shown in Table~\ref{ablation}, omitting preprocessing causes a steep drop in TraceRAG's performance: only 2 out of 7 queries correctly return the relevant malicious behaviors (Q1: access or collection of sensitive user data; Q4: use of obfuscation or encryption). Moreover, even for those two correctly retrieved cases, the subsequent analysis reports are relatively superficial, and call chains are rarely reconstructed successfully.

\begin{table}[htbp]
\centering
\caption{Comparison of Detection Results and number of codes analyzed}
\label{ablation}
\resizebox{\linewidth}{!}{%
\begin{tabular}{lcccccc}
\toprule
\multirow{2}{*}{} & \multicolumn{2}{c}{\textbf{TraceRAG - No Split and Cleaning}} & \multicolumn{2}{c}{\textbf{TraceRAG}} & \multicolumn{2}{c}{\textbf{VT}} \\
\cmidrule(r){2-3} \cmidrule(r){4-5} \cmidrule(r){6-7}
 & Detection & Codes & Detection & Codes & Detection & \\
 &           & Analyzed &           & Analyzed &           & \\
\midrule
Q1  & 1 & 2 & 1 & 3 & 1 & \\
Q2  & 0 & 3 & 0 & 3 & 0 & \\
Q3  & 0 & 1 & 1 & 4 & 1 & \\
Q4  & 1 & 2 & 1 & 3 & 1 & \\
Q5  & 0 & 1 & 0 & 1 & 0 & \\
Q6  & 0 & 3 & 0 & 3 & 0 & \\
Q7  & 0 & 1 & 0 & 1 & 0 & \\
Q8  & 0 & 3 & 1 & 3 & 1 & \\
Q9  & 0 & 1 & 1 & 4 & 1 & \\
Q10 & 0 & 4 & 1 & 4 & 1 & \\
Q11 & 0 & 1 & 1 & 1 & 1 & \\
\midrule
\textbf{Total} & \textbf{2} & \textbf{22} & \textbf{7} & \textbf{30} & \textbf{7} & \\
\bottomrule
\end{tabular}%
}
\end{table}

A meticulous post‐hoc review reveals that these results are mainly caused by two reasons. First, without splitting and cleaning, the top-5 snippets retrieved for most queries bear no relation to the corresponding behavior and are subsequently discarded by the Relevance-Reviewer. As a result, LLM-Analyzer fails to receive any relevant code, causing the malicious behavior to be overlooked. Second, even when relevant snippets are retrieved, the input remains overly large and contains extraneous code, which degrades analysis quality. Moreover, follow-up queries frequently target incorrect snippets, causing the analysis process to terminate prematurely.

These findings demonstrate that code splitting and cleaning process not only improve retrieval accuracy but also enhance analysis quality. Shorter and cleaner inputs allow TraceRAG to focus on simpler tasks, which also reduces hallucination issues.

\subsubsection{Influence of Multi-turn Conversation}
Android applications routinely invoke functionality across multiple classes, and malware authors often employ dynamic loading or obfuscation to conceal their intent. As a result, key behaviors may only become apparent when the analyzer can follow these multi‐step invocation paths. To support this, TraceRAG allows the LLM-analyzer to generate follow‐up queries for additional code snippets, iteratively piecing together the full execution chain and draw accurate conclusions. To evaluate the necessity of this multi-turn analyze, we remove the follow-up query part from LLM-Analyzer's prompt, only allow it to reason in a single pass. The results mirror those of the preprocessing ablation, TraceRAG produces only superficial summaries, and cannot reconstruct call chains. It can only describe possible intent or usage of the code, but cannot explain how those intents are achieved or what outcomes they produce. These findings underscore the importance of multi‐turn interaction, as successive queries enable TraceRAG to derive deeper insights into complex malicious behaviors.

\subsection{RQ3: Specialist Feedback and Usability Study}
Since TraceRAG focuses on behavior-level detection and interpretability, rather than merely performing binary malware-versus-benign classification, it fundamentally differs in objective and output format from existing Android malware detection methods. As a result, direct quantitative comparisons with conventional detection systems are not entirely meaningful. To address this, we conduct a structured developer study aimed at evaluating the practical utility of our system. The goal is to evaluate how well the system helps developers understand Android applications, detect malicious behavior, and reduce analysis time.

\subsubsection{Participants and Survey Design}

We invite specialists from the Google Android Security Team to review a set of analysis reports generated for Android applications that are flagged as malware on AndroZoo. A total of 42 reports are shared, and the specialists provide 31 pieces of feedback covering 24 APKs, selected at random. For APK reports evaluated by multiple specialists, we use the average of their scores.

The survey includes three main sections. The first section evaluates the \textit{usefulness of each component} of the LLM-generated report, including the App Info, Overall Summary, Detailed Analyses, and Conclusion. The second section focuses on the system’s \textit{ability to identify malicious behaviors}, asking participants to assess its performance across predefined behavior categories such as information theft, monetary abuse, and privilege escalation. The third section covers \textit{overall accuracy and findings}, including whether the report correctly identifies malicious functions, whether participants discover new insights, and how confident they feel in the report’s conclusions.

The survey uses a mix of 5-point Likert scale questions, multiple-choice questions, and a few open-ended items (in the general survey only). We release the survey along with the source code.

\subsubsection{Feedback Results}
We analyze the collected report from the following three perspectives.

\begin{figure}
\centering
\includegraphics[width=0.9\linewidth]{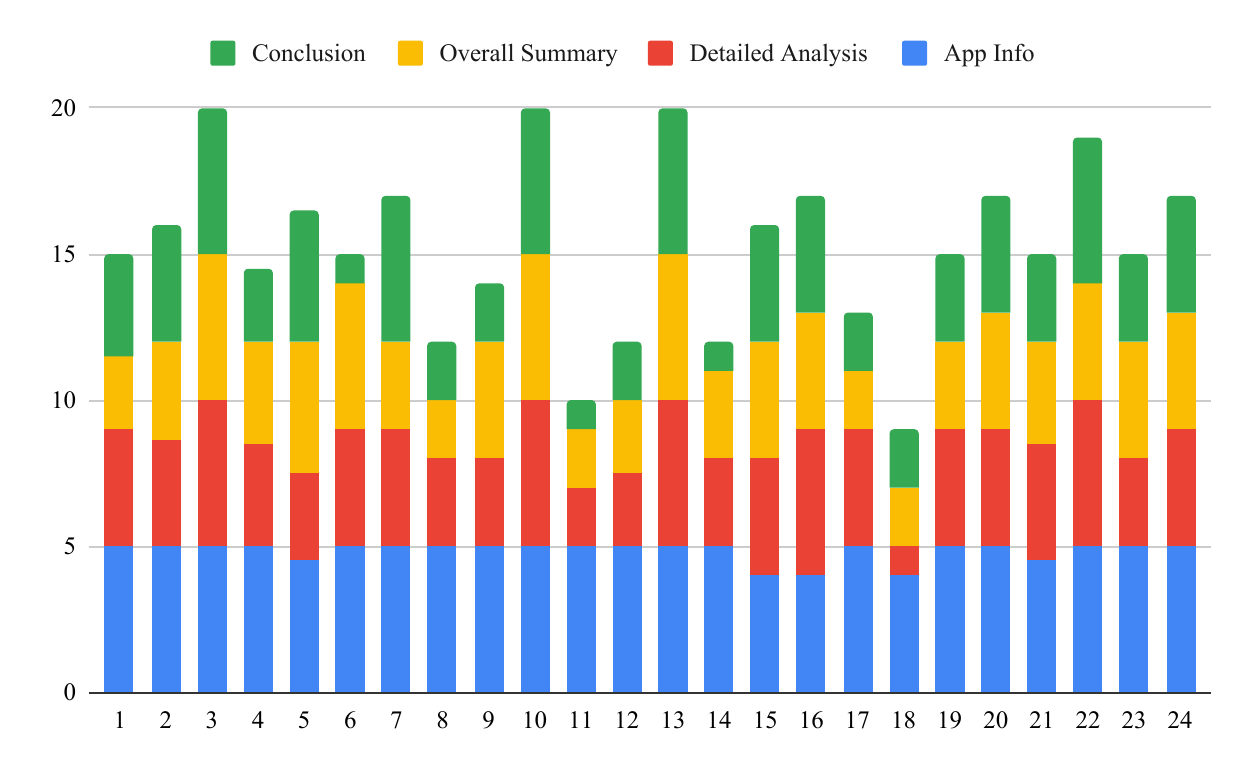}
\caption{Usefulness of Report Components}
\label{fig:section_usefulness}
\end{figure}

\noindent\textbf{Usefulness of Report Components.}
Participants rate the usefulness of four key components of the LLM-generated report: App Info, Overall Summary, Detailed Analyses, and Final Conclusion. The App Info section receives the highest average usefulness score of 4.83, reflecting consistent clarity and relevance. The Detailed Analyses and Overall Summary also received positive ratings, with average scores of 3.69 and 3.53, respectively. Several participants mention that these sections highlight suspicious behaviors and reduce the amount of manual inspection needed. The Summary section shows more variation in ratings, with an average score of 3.23, suggesting that further refinement may improve focus and clarity. Figure~\ref{fig:section_usefulness} shows the distribution of scores for each section.

\noindent\textbf{Malicious Behavior Identification.}
To assess how effectively the system identifies malicious behaviors, participants report whether the flagged behaviors in each report align with their own analysis. Out of 29 responses, 65.52\% indicate alignment, 20.69\% disagree, and 13.79\% remain unsure. These responses suggest the system generally captures relevant behavior patterns, though precision can still improve.

Participants submitted six reports of incorrect behavior categories across 31 reports. Three relate to \textit{Information Theft and Abuse}, two to \textit{Privilege Abuse and System Exploitation}, and one to \textit{Monetary Fraud and Financial Abuse}. This shows that misclassifications occur across all behavior types, with slightly more in the information theft category.

When asked whether any major behavior types are missed, 73.33\% of 30 responses say "no," 6.67\% say "yes," and 20\% remain unsure. The two participants who believe behaviors are missed point to indicators related to \textit{Privilege Abuse and System Exploitation}. This indicates that, while the system broadly covers key behaviors, privilege-related cases may need better detection.

\noindent\textbf{Accuracy and Findings.}
Participants evaluate the overall accuracy of the system’s conclusions. Out of 30 responses, 53.33\% state the final app classification (malware or not) is accurate, 16.67\% say it is inaccurate, and 30\% are unsure. This suggests a general sense of reliability.

Among 25 responses related to confirmed malware samples, 56\% say the system correctly identifies malicious methods or functions. This supports the system's ability to highlight critical behaviors. However, 16\% report missing key methods, and 28\% are unsure, indicating a chance to improve detection—especially for behaviors that are subtle, hidden, or depend on context.

When asked whether flagged methods and classes are helpful for malware detection, 34.48\% say all are useful, another 34.48\% say most are useful but some are misleading, 6.9\% say only a few are useful, and 6.9\% say most are not useful. Free-text responses mention concerns such as too much detail, irrelevant SDK-related content, and missing context like Android version differences or dynamically loaded code.

Moreover, most participants express confidence in their ability to evaluate the system. Out of 31 responses, 15 rate their confidence at 4 out of 5, and 9 give a full 5. Only 7 rate below 4.
These findings highlight that the system is capable of identifying malicious behavior across a range of samples. They also point to an opportunity to further align flagged content with user expectations by refining how suspicious or malicious activity is prioritized and presented.
\section{Discussion}

In this section, we discuss the current limitations in TraceRAG and outline possible directions for improvement.

\subsubsection{Information Loss during Code Cleaning}
TraceRAG’s preprocessing pipeline removes dead or highly obfuscated code to reduce noise, but this step can inadvertently eliminate legitimately executed logic. As a result, the LLM-Describer may generate descriptions that omit critical behavior, which in turn impacts downstream analysis. Our ablation study also reveals the inverse problem: skipping the cleaning step leads to generated descriptions containing useless information from dead code, which degrades the quality of further analysis. This issue only arises occasionally when input code is extremely obfuscated, so we have to trade off the conciseness and completeness of the code. Despite extensive prompt tuning to mitigate information loss, some essential logic may still occasionally be removed. In future work, we plan to evaluate more advanced LLMs and develop adaptive cleaning strategies that balance noise reduction with preservation of critical code paths.

\subsubsection{Limitations of Java-only Analysis}
While static analysis of Java code uncovers many malicious behaviors, TraceRAG currently misses functionality implemented outside of the indexed app code. Some malware hides core logic in native libraries—exposed only as native method stubs and hidden in binaries that standard decompilers cannot inspect. At the same time, follow-up queries generated by LLM-Analyzer sometimes refer to Android SDK methods whose implementations lie outside our indexed corpus, preventing TraceRAG from retrieving or examining their code. Although LLMs possess strong general reasoning capabilities, they may struggle with highly specialized, domain-specific tasks. In future work, we plan to incorporate such considerations into our framework to further improve TraceRAG’s robustness.

\subsubsection{Compute Resources Constraints}
All LLMs used in our experiments are based on the state-of-the-art OpenAI o3-mini model, which provides TraceRAG with strong reasoning capabilities under an acceptable cost constraint. However, due to limitations in computational resources and budget, we are unable to conduct experiments on a larger scale. While local deployment of LLMs would allow for greater flexibility, it requires significantly more hardware support. After extensive testing, we opted to use OpenAI's API for all experiments. Nevertheless, the API imposes restrictions on traffic load, preventing us from utilizing more threads to accelerate processing. As a result, we have to make trade-offs between experimental time cost and dataset scale.
\section{Conclusion}
In this work, we present TraceRAG, a novel framework that integrates LLMs with RAG to enable explainable Android malware detection and analysis. We first process each app into cleaned, method-level code snippets and generate concise semantic summaries for efficient indexing; when a specific behavior query is issued, TraceRAG retrieves the most relevant snippets, performs iterative analysis to identify malicious behaviors and corresponding code paths, and then compiles the findings into a structured, human-readable report. Grounded by comprehensive evaluations, TraceRAG demonstrates strong performance in both malware detection and behavior identification. Compared to existing malware detection platforms that provide only simple behavioral logs or isolated code fragments, TraceRAG-generated reports are more traceable, clearly organized, and instructive. A promising direction for future work would be to extend the LLM-driven analysis framework to support the examination of native libraries and dynamically loaded components, thereby achieving end-to-end coverage of all execution paths.
\section*{Acknowledgements}
This research is supported by the Commonwealth Cyber Initiative (CCI)’s Coastal Virginia Node and Northern Virginia Node (Grant ID: 765711). The authors also thank the Google Android Security Team for providing valuable feedback on this work.

\clearpage
\newpage
\bibliographystyle{IEEEtran}
\bibliography{main}

\end{document}